# Detection of magnetic nanoparticles (MNPs) using spin current nano-oscillator (SCNO) biosensor: A frequency-based rapid, ultra-sensitive, magnetic bioassay

Renata Saha[1], Kai Wu[1, *], Diqing Su[2], and Jian-Ping Wang[1, *]

[1]Department of Electrical and Computer Engineering, University of Minnesota, Minneapolis, Minnesota 55455, USA

[2]Department of Chemical Engineering and Material Science, University of Minnesota, Minneapolis, Minnesota 55455, USA

*Corresponding author E-mails: wuxx0803@umn.edu (K. W.) and jpwang@umn.edu (J.-P. W.)

This Letter is a micromagnetic simulation-based study on the GHz-frequency ferromagnetic resonances for the detection of magnetic nanoparticles (MNPs) using spin current nano-oscillator (SCNO) operating in precession mode as a spintronic biosensor. The magnetic stray fields from the MNPs in an antibody-antigen-MNP complex on the SCNO surface modify the ferromagnetic resonance peaks and generate measurable resonance peak shifts. Moreover, our results strongly indicate the position-sensitive behavior of the SCNO biosensor and ways to eradicate this effect to facilitate better bio-sensing performance. Additionally, a study has been made on how nanoparticles with different sizes can alter the SCNO device performance. This simulation-based study on the SCNO device shows a promise of frequency-based nano-biosensor with a sensitivity of detecting even a single MNP, even in presence of thermal noise.

   It has been two decades since Baselt *et. al*[1] has designed the Bead Array Counter (BARC) that first showed the experimental possibility of bio-detection for multilayer giant magnetoresistance (GMR) with MNPs as biomarkers. Ever since then, magnetic biosensing for point-of-care (POC) detection of diseases using magneto-resistive (MR) sensors[2–13] have been explored intensively and has been subjected to extensive reviews[14–16]. Sandwich-based bioassay[9,17], flow cytometry[18,19] and microfluidic channel[20] are the most common techniques for magnetic biosensing. The most attractive part about biosensing with spintronic sensors lies in the fact that biomedical samples exhibit negligible magnetic background which suppresses noise from cellular matrix to a great extent[21]. These spin-valve sensors have the sensitivity of detecting miniscule change in magnetic field even from those of surface functionalized MNPs[9,11] or from a single magnetic bead[22,23] by translating the presence of MNP(s) to cause a variation in the static magnetic configuration of the sensor's active sensing layer. This alters the device resistance which is manifested and measured as the change in voltage.

   However, these MR sensors tend to suffer from high background noise levels at room temperature performance which causes to compromise the sensitivity of the device. Thus, recently, frequency-based approach for detection of magnetic field[24–27] and MNPs[28–30] have been implemented through micromagnetic simulations and validated through experimental analysis. However, majority of them were using magnonic crystals[24,29], ferromagnetic nanodots[28] and/or nanodiscs[30]. The ferromagnetic resonance (FMR) frequency of the device interacts directly with the stray field of the MNP[31] and thereby causes a shift in the peak frequency of the device. The shift in the peak frequency has been experimentally demonstrated to depend on the concentration of the MNPs[28], the size of the MNPs[28,30] and even the position of the MNPs[30] on the magnonic crystal surface. The main advantage of a frequency-based, dynamic approach over the static MR-based sensing is that the device response is linear over a large range of the externally applied magnetic field leading to more prominent frequency shifts. This frequency is typically of the orders of several GHz, way too higher compared to the low frequency1/f noise, and hence devoid of DC voltage-level drift.

In this regard, spin torque nano-oscillators (STNO)[32–36] driven by spin transfer torque (STT)[37,38] deserve special mention for frequency-based magnetic biosensing[39,40]. The stack structure of these STNO is similar to a spin-valve structure that is composed of a nonmagnetic metallic layer sandwiched between two ferromagnetic layers. The spin-polarized electric current passing through a thin ferromagnetic layer dynamically excites the magnetic moment of that layer through a transfer of spin angular momentum. However, STNO devices are limited by consumption of large currents as the electrons are limited by a total angular momentum of $\hbar/2$[41]. Even more, larger the magnetic moment, larger is the current required to operate. But the trade-off is that it yields larger thermal stability with larger magnetic moment. In this respect, in-plane magnetized spin Hall nano-oscillators (SHNO)[41–48] consisting of a heavy metal (HM)/ferromagnetic metal (FM) stack structure do not have that limitation of angular momentum as constant scattering takes place at HM/FM interface. Besides, no electron is required to flow through the active FM. Consequently, unpredicted damage due to electromigration and ohmic heating is prevented in spin Hall effect (SHE) devices. Unlike STT devices, SHE devices support magneto-optical measurements in direct contact with the active area of the device. Even more, from fabrication point-of-view, spin Hall nano-oscillators (SHNO) are easier to fabricate.

There are two conditions to trigger self-oscillations in SHNO devices: first, the dynamic damping must be completely balanced by spin current; second, the current to balance the spin torque and the damping torque in the self-oscillation state should be larger than the critical current to destabilize the initial state. Since the first condition is not satisfied, SHNOs with perpendicular magnetized anisotropy (PMA) are not feasible as was proved theoretically by Tomohiro Taniguchi[49]. The spin-orbit effects from the HM/PMA-FM bilayer system that contribute to the current-induced phenomena including the spin Hall effect, the Rashba effect[50], and the Dzyaloshinskii-Moriya interaction (DMI)[51], all of which originate from the broken inversion symmetry at the HM/PMA-FM interface. A combination of all these interactions contribute to a far more stable oscillator system in comparison to in-plane SHNO systems. It is this possibility to induce dynamical states of HM/FM bilayer or alter their static configuration by the current-induced spin torque (ST)[37,38] has triggered extensive experimental and theoretical research. In this respect of operation, one of the promising devices is the PMA-spin current nano-oscillators (SCNO)[52] devices. However, the detailed investigation of SCNOs in terms of frequency-based nano-biosensors have not been made.

As per our best knowledge, for the first time, we investigate the feasibility of spin current nano-oscillator (SCNO) device as a frequency-based spintronic biosensor through micromagnetic simulations on Mumax3[53]. The SCNO device has been simulated numerically by solving the Landau–Lifshitz–Gilbert (LLG) equation (1) in addition to a spin orbit torque (SOT):

$$\frac{d\mathbf{m}}{dt} = \gamma_0 \mathbf{h}_{eff} \times \mathbf{m} + \alpha \mathbf{m} \times \frac{d\mathbf{m}}{dt} + \frac{u}{t}\mathbf{m} \times (\mathbf{m}_p \times \mathbf{m}), \qquad (1)$$

where, $\mathbf{m} = \mathbf{M}/M_s$ is the normalized magnetization, $\gamma_0 = 1.85 \times 10^{11}$ rad T$^{-1}$s$^{-1}$ is the gyromagnetic ratio, $\mathbf{h}_{eff} = \mathbf{H}_{eff}/M_s$ is the reduced effective field, $t$ is the thickness of the ferromagnetic (FM) layer, $\mathbf{m}_p$ is the current polarization vector, $u = \gamma_0(\frac{\hbar jP}{2eM_s})$, and $j$ is the density of the spin current. The values of parameters $M_s$, P, $\alpha$ along with dimensions of the magnetic thin film are listed in Table 1. All parameters that define the FM layer are adopted from Ref.[54] and that to define the MNP are adopted from Ref[55].

Table I. Micromagnetic simulation parameters for SCNO biosensor

| Parameters | Description | Values |
|---|---|---|
| **FM nanopillar Dimension** | Length × Width × Thickness | 160 nm × 80 nm × 5 nm |
| **Cell Size** | Length × Width × Thickness | 2.5 nm × 2.5 nm × 5 nm |
| $\alpha$ | Gilbert damping factor | 0.015 |
| A | Exchange constant | $13 \times 10^{-12}$ J/m |
| P | Spin Hall Angle | 0.6 |
| $M_s$ | Saturation magnetization | $1200 \times 10^3$ A/m |
| $Ku_1$ | First order uniaxial anisotropy constant | $0.7 \times 10^{-6}$ Jm$^{-3}$ |
| DMI | Dzyaloshinskii-Moriya interaction | $0.7 \times 10^{-4}$ Jm$^{-2}$ |
| $\mu_0$ | Permeability of free space | $4\pi \times 10^{-7}$ WbA$^{-1}$m$^{-1}$ |

For a bilayer of PMA ferromagnet (FM) and heavy metal (HM), under an externally applied current and uniform DC magnetic field, the device operates in precession mode. It is with a precession frequency that the bare SCNO device oscillates (see Supplementary Movie SM1), referred to in this Letter as the 'peak frequency' or in other words the frequency which has the maximum intensity. We have demonstrated how the peak frequency shifts with respect to regularly and/or randomly spaced single and/or a cluster of MNP(s). In addition, how different sized MNPs can affect the peak frequency shift of the SCNO device have been investigated (see Supplementary Movie SM2 & SM3). Finally, discussions follow on how we can optimize the SCNO device performance for it to be best fit in magnetic biosensing application.

Figure 1(a) gives a schematic view of the SCNO biosensor array with the ferromagnetic nanopillar of dimensions 160 nm × 80 nm × 5 nm located 0.5 μm apart such that the stray fields of adjacent PMA-FM nanopillars do not influence the device performance (see Supplementary

Information S3). Figure 1(a) also shows the mechanism in which the SCNO biosensor would facilitate magnetic biosensing through formation of target antibody-antigen-MNP complex. The magnetic material parameters to define the SCNO biosensor in micromagnetic simulations are listed in Table I. The properties of the MNP used in this simulation work are specified in Table II. Figure 1(b) is a zoomed in image of a single FM nanopillar. On passing a charge current ($J_c$, A/m$^2$) through the HM layer along **-x** direction, it causes spin accumulation along **±y** and generation of a spin current along **z** direction (see Figure 1(b)). When a magnetic field, $H_{dc}$ (in Oe) is externally applied along **+y** direction, the spin current causes the FM nanopillar to operate in precession. The effects of a reversed direction of $J_c$ and $H_{dc}$ on SCNO device performance have been explored in Supplementary Information S2. The color and symbol codes for antigen, antibody, MNP, target antibody – antigen – MNP complex, FM layer, HM layer & substrate of the SCNO biosensor used throughout the figures in this Letter are specified in a separate column in Figure 1. The performance of the designed SCNO biosensor in this Letter has been reported at T = 0 K. However, the arguments concerning the thermal effects on its performance have been made in Supplementary Information S1.

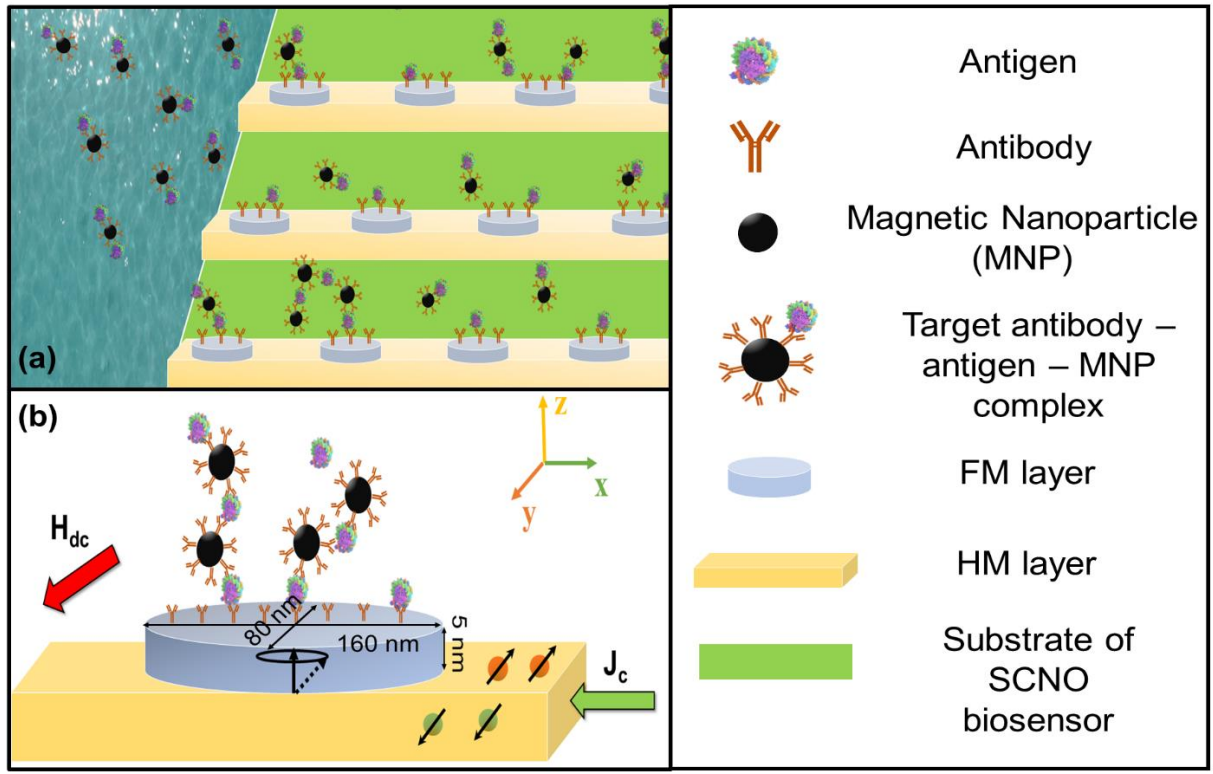

Figure 1. (a) Schematic of the spin current nano-oscillator (SCNO) biosensor array with the mechanism of target antibody - magnetic nanoparticle (MNP) - antigen complex demonstrated. (b) A single 160 nm × 80 nm × 5 nm, perpendicularly magnetized (PMA) ferromagnetic (FM) nanopillar of the SCNO biosensor array zoomed in. The charge current density ($J_c$) to the heavy metal (HM) layer is along **-x** direction and the externally applied magnetic field ($H_{dc}$) is directed

along **+y** direction. The black-dashed arrow in the ferromagnetic nanopillar demonstrates precession mode operation of the device.

Table II. Micromagnetic simulation parameters for MNP(s)

| Parameters | Description | Values |
|---|---|---|
| $\alpha$ | Gilbert damping factor | 0.1 |
| A | Exchange constant | $2.64 \times 10^{-11}$ J/m |
| P | Spin Hall Angle | 0.6 |
| $M_s$ | Saturation magnetization | $3.5 \times 10^5$ A/m |
| $Ku_1$ | First order uniaxial anisotropy constant | $1.25 \times 10^4$ Jm$^{-3}$ |

For an externally applied magnetic field, **H**$_{dc}$ = 1.1 kOe in Figure 2(a) the peak frequency for the SCNO biosensor changes with variation of externally applied current through the heavy metal (i, mA). As reported in previous literatures for in-plane spin Hall nano-osillators (SHNO)[41,42], with increase in current for a constant magnetic field, the peak frequency (in GHz) decreases while the intensity of the main frequency component of the SCNO device (in arbitrary units (a.u.)) increases as is expressed in Figure 2(b). Again, for an externally applied current of i = 15 mA (current density, $1.5 \times 10^8$ A/cm$^2$) for varying DC magnetic field, we observe a clear shift in the peak frequency value in Figure 2(c). The black diamonds represent the calculated ferromagnetic resonance frequency (f$_{FMR}$) value for a particular externally applied magnetic field as calculated by the normal magnetization Kittel Equation, $f_{FMR} = \gamma(H_{dc} - 4\pi M_s) + (4\pi\gamma M_s)$. As reported earlier[52], the precession frequency for the designed SCNO biosensor always lies below the calculated FMR frequency. The trajectory of the magnetization vectors (**m**$_x$, **m**$_y$ and **m**$_z$) due to precessional motion of the FM nanopillar has been demonstrated in Figure 2(d). Discussions relating power consumption and device performance have been made in Supplementary Information S5.

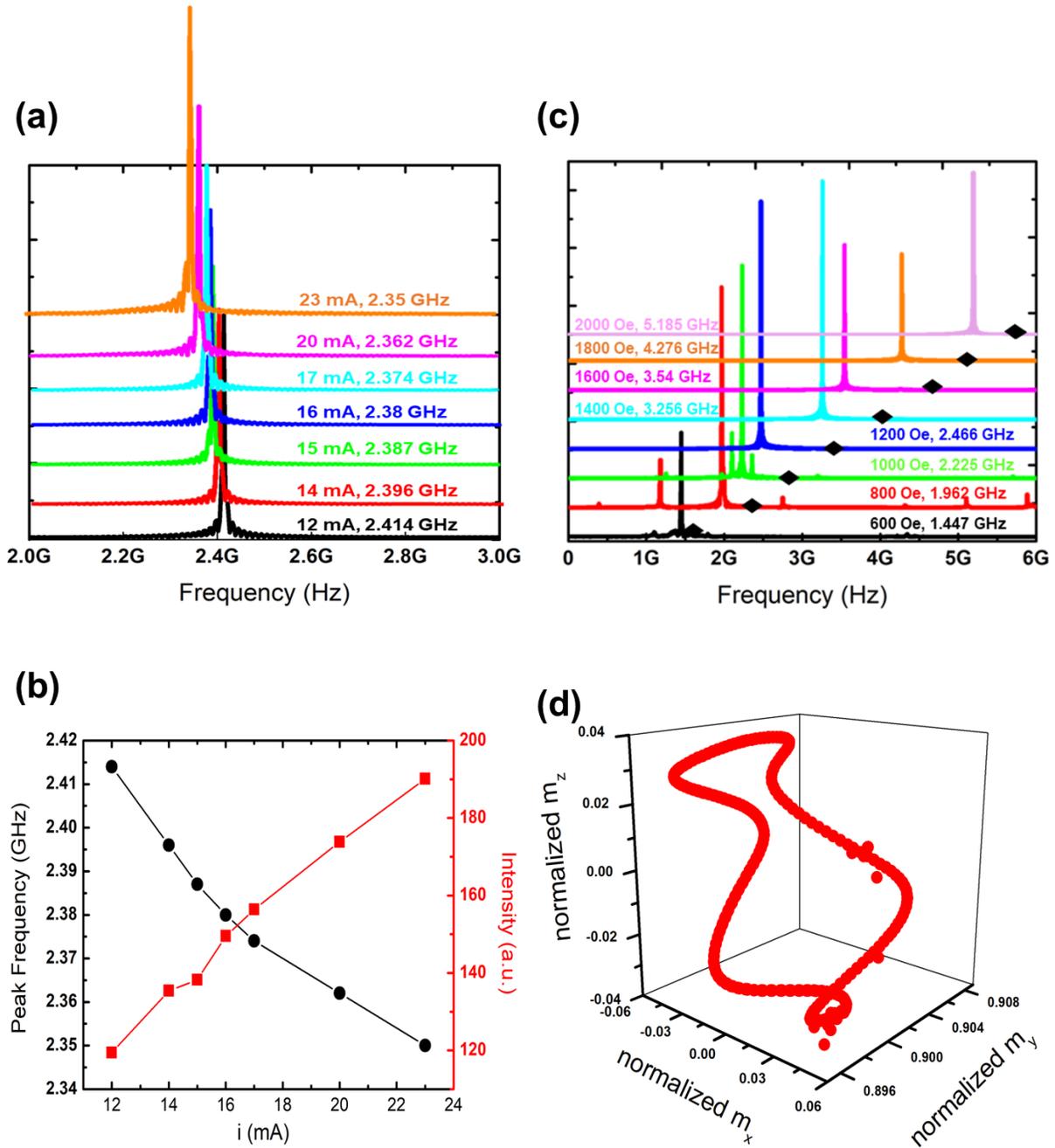

Figure 2. (a) Variation of peak frequency shift of a SCNO biosensor with current (i, mA) at $H_{dc}$ = 1.1 kOe. (b) As a follow-up from part (a), representation of the variation of peak intensity and magnitude of peak frequency with current (i, mA). (c) Variation of the peak frequency with applied magnetic field ($H_{dc}$, Oe) at i = 15 mA. The black diamonds represent the value of the ferromagnetic resonant (FMR) frequency at the corresponding frequency calculated by the Kittel Equation. (d) The magnetization vector (normalized values $m_x$, $m_y$ and $m_z$) trajectory due to their precessional motion due to i = 15 mA and $H_{dc}$ = 1.1 kOe.

In Figure 3, a SCNO device devoid of any MNP positioned on the sensor surface has been referred to as the 'bare' SCNO device. Figure 3(a) shows 5 independent positions of a single MNP of 20 nm diameter on a SCNO biosensor. The 5 independent positions correspond to the following co-ordinates: (i) = (0, 0), center of the SCNO device; (ii) = (40, 20), first quadrant; (iii) = (-40, 20), second quadrant; (iv) = (-40, -20), third quadrant; (v) = (40, -20), fourth quadrant. Figure 3(b) correspond to 4 independent positions of 6 MNPs, each of 20 nm in diameter and the center of each MNP spaced regularly at 30 nm apart from each other. The 4 independent positions correspond to the following co-ordinates: (i) = first quadrant; (ii) = second quadrant; (iii) = third quadrant; (v) = fourth quadrant. Figure 3(c) corresponds to 8 & 10 MNPs positioned at the center of the SCNO biosensor, each of 20 nm in diameter and center of each MNP regularly spaced at 30 nm from each other.

Corresponding to the highlighted picture background color-codes, the peak frequencies for cases in Figure 3(a), (b) & (c) have been displayed in Figure 3(d), (e) & (f), respectively. With respect to a bare SCNO biosensor, Figure 3(d) shows peak frequency shift for the 5-independent positions of a single 20 nm MNP while Figure 3(e) shows peak frequency shift for the 4-independent positions of a six, 20 nm MNPs with their centers separated by 30 nm distance. Both Figure 3(d) & (e) show that the peak frequency varies for different positions of the MNPs. Furthermore, for both the cases of a single MNP and for the case of 6 MNPs, the peak frequency for the positions in the first & fourth quadrant and for the positions in second & third quadrant are same. Therefore, one can conclude that the two halves of the SCNO device work differently due to unique magnetization distribution (see Supplementary Information S4). Figure 3(e) demonstrates the variation in peak frequency due to presence of 8 and 10 MNPs on the SCNO biosensor surface with respect to a bare SCNO device. In summary, Figure 3(a)-(f) validates the fact that SCNO biosensor performance is position specific, precisely, the two identical halves of the device work uniquely. As much as this position sensitivity of the designed SCNO biosensor is detrimental to biosensing performance for magnetic biosensors, one cannot deny the fact that the other magnetic biosensors, including the most celebrated biosensor in magnetic biosensing, GMRs are position sensitive too. For instance, analytical studies by Klein *et. al*[56] had shown how the edges of the GMR sensors are more sensitive towards detection of MNPs than the remaining part of the sensor.

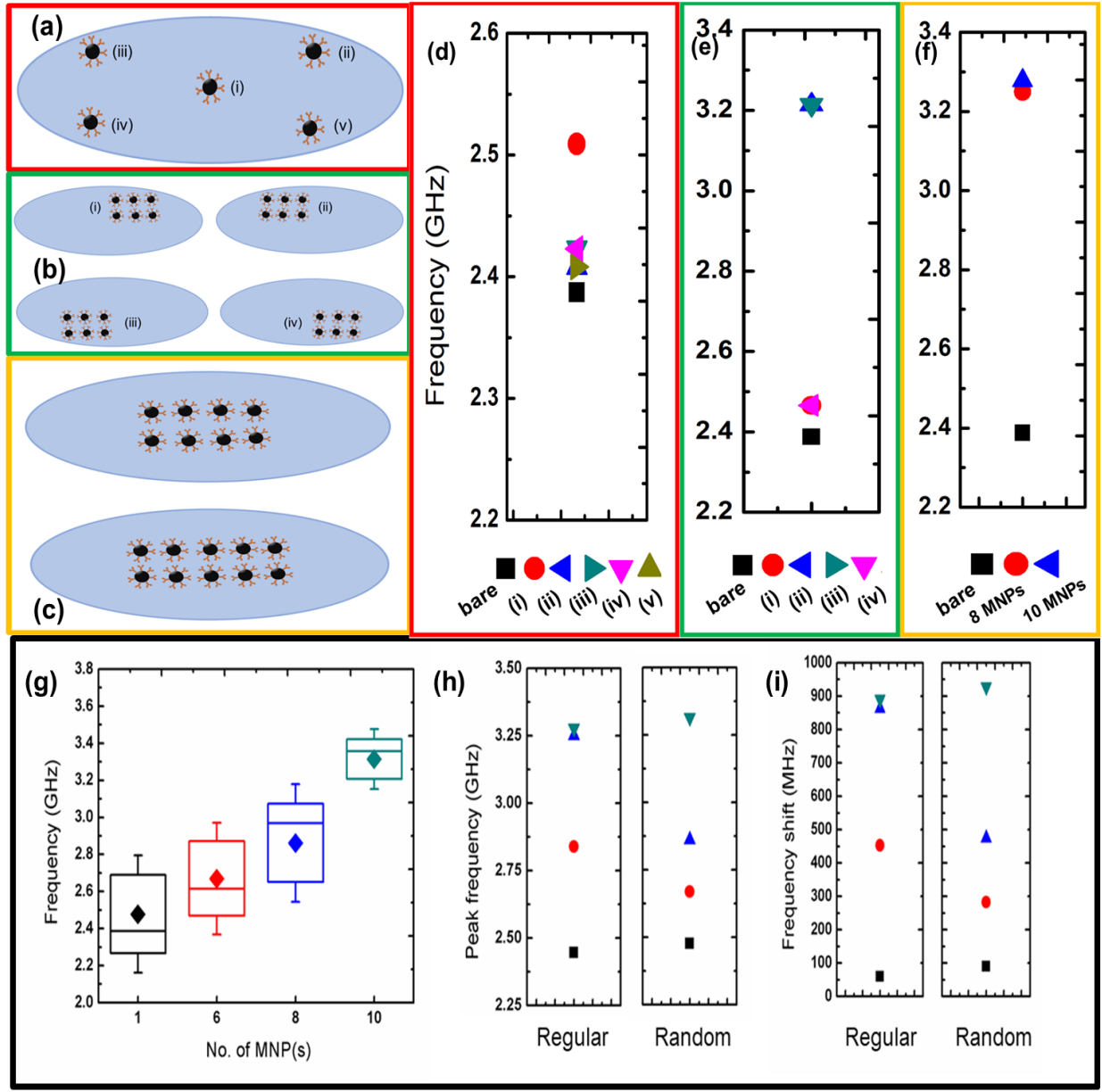

Figure 3. Demonstration of positional sensitivity along with MNP detection capability of the SCNO biosensor. Top view of the FM nanopillar containing (a) one MNP at 5 different positions on the SCNO biosensor marked (i), (ii), (iii), (iv) & (v). Each of the 5 cases are independent of each other; (b) 6 MNPs at 4 different positions on the SCNO biosensor marked (i), (ii), (iii) & (iv). Each of the 4 cases are independent of each other; (c) 8 MNPs & 10 MNPs positioned at the center of the SCNO biosensor. In all the cases (a)-(c), each of the MNPs were identical of 20 nm in diameter and the distance between centers being 30 nm. With respect to a bare SCNO device, (d) Peak frequency for the conditions of MNP position demonstrated in (a). (e) Peak frequencies for the conditions of MNP position demonstrated in (b). (f) Peak frequencies for the conditions of MNP position demonstrated in (c). (g) Variation of the peak frequencies with 1, 6, 8 & 10 no. of MNPs of 20 nm diameter situated at random, uncontrolled positions on the SCNO biosensor. (h)

Comparison between the peak frequency (in GHz) variation for regular and random position of MNP on the biosensor. (i) The peak frequency shift (in MHz) for values in (h) from that of the bare SCNO device.

However, in real platforms for magnetic biosensors, the MNPs are unlikely to be uniformly spaced as was the case in Figure 3(a)-(c). To demonstrate a more realistic SCNO performance, we have further investigated the cases of 1, 6, 8 and 10 MNPs, but this time for 5 random arrangements on the SCNO surface for each of the 4 number of MNPs. Figure 3(g) demonstrates the box-whisker plots for the peak frequencies of 5 cases of randomly positioned 1, 6, 8 & 10 MNPs, each. The mean values for 1, 6, 8 & 10 MNPs have been found to be 2.378 GHz, 2.556 GHz, 2.862 GHz and 3.314 GHz respectively and are symbolized in varied colored diamonds in Figure 3(g). Furthermore, as the number of MNPs on the SCNO biosensor surface increases, the deviations decreases, that is, the length of the box in the box-whisker plot decreases. Figure 3(h) shows the comparison of peak frequencies of regularly spaced 1, 6, 8 & 10 MNPs as discussed in Figure 3(a)-(f) to that of the mean peak frequency values for randomly spaced 1, 6, 8 & 10 MNPs as discussed in Figure 3(g). From Figure 3(g), it is evident that the trend for both regularly and randomly spaced MNPs are the same, which is, the peak frequency (in GHz) value increases with increase in number of the MNPs from 1, 6, 8 to 10 MNPs. This fact is validated from Figure 3(h) by the shift in peak frequency (in MHz) from the bare SCNO device. The demonstration of the cases for uniformly spaced MNPs validates the fact that an SCNO biosensor is position sensitive. The similar trend in the peak frequency shifts between randomly spaced MNPs and regularly spaced MNPs draws a more realistic picture towards SCNO biosensor performance because in real experiments, the MNPs would be quiet randomly positioned.

In Figure 4, we have defined a single MNP of 7 different diameters at random positions on the SCNO device surface. For 6 random positions on the SCNO device surface, MNPs of diameters, 10 nm, 20 nm, 25 nm, 30 nm, 35 nm, 40 nm and 45 nm show a mean peak frequencies 2.418 GHz, 2.43 GHz, 2.514 GHz, 2.554 GHz, 2.562 GHz, 2.647 GHz and 2.56 GHz, respectively. With increase in the diameter of the MNPs, the mean peak frequency increases gradually until at 45 nm where a sudden drop in mean peak frequency is observed. Analogous results, in terms of the GMR signal level, were experimentally observed by Wang *et. al*[57] for the purpose of GMR biosensing. In comparison between large and small size of MNPs, the latter encourage greater degree of Brownian motion which in turn facilitates greater diffusion and binding capacity of MNPs with the SCNO sensor surface. Therefore, with increased diameter of the MNP, the binding tendency to the SCNO sensor surface decreases significantly thereby leading to a decrease in peak frequency.

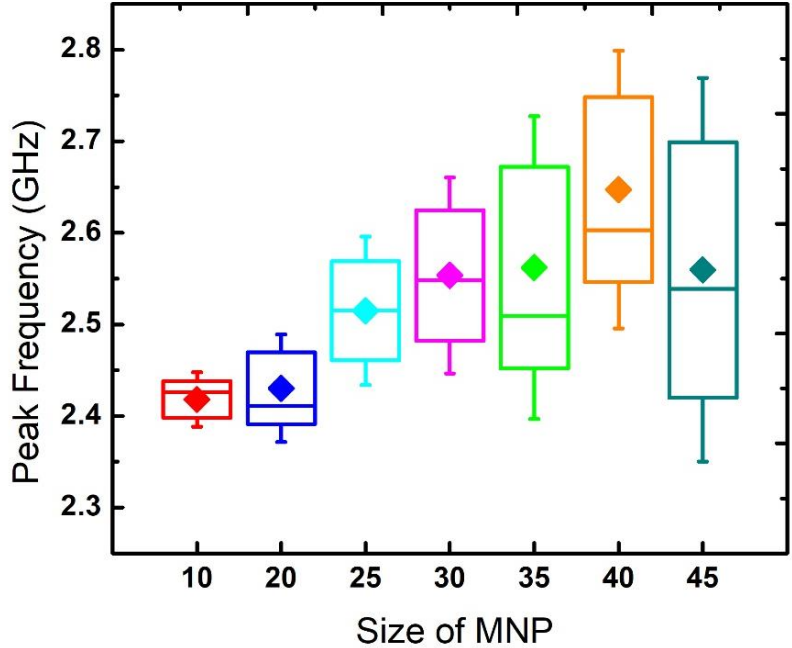

Figure 4. SCNO performance for varied sizes of a single MNP at random positions on the sensor surface.

In conclusion, we have proposed and investigated the feasibility of a spin current nano-oscillator (SCNO) biosensor with perpendicular magnetic anisotropy (PMA) as a frequency-based biosensor. Through micromagnetic simulations, we have demonstrated that the SCNO biosensor has the sensitivity of detecting even a single MNP and its performance varies with the position of the MNP. That the performance of a SCNO biosensor varies with the position of the MNP(s). However, in real experiments, with random position of the MNP(s), the position specific behavior of the SCNO biosensor can be eliminated to a great extent. Unlike MR sensors, the SCNO biosensor performance is not noisy at room temperature yielding a more realistic device performance. Finally, in order to observe a distinct peak shift on addition of MNPs, a standard binding process is required to be initiated. Therefore, optimizing the size of the MNPs to facilitate binding to observe a clear shift of frequency is extremely essential.

This study was financially supported by the Institute of Engineering in Medicine of the University of Minnesota through FY18 IEM Seed Grant Funding Program, the National Science Foundation MRSEC facility program, the Distinguished McKnight University Professorship, the Centennial Chair Professorship, and the Robert F Hartmann Endowed Chair from the University of Minnesota.

# Supplemental Information

**Detection of magnetic nanoparticles (MNPs) using spin current nano-oscillator (SCNO) biosensor: A frequency-based rapid, ultra-sensitive, magnetic bioassay**

Renata Saha[1], Kai Wu[1, *], Diqing Su[2], and Jian-Ping Wang[1, *]

[1]Department of Electrical and Computer Engineering, University of Minnesota, Minneapolis, Minnesota 55455, USA

[2]Department of Chemical Engineering and Material Science, University of Minnesota, Minneapolis, Minnesota 55455, USA

*Corresponding author E-mails: wuxx0803@umn.edu (K. W.) and jpwang@umn.edu (J.-P. W.)

**Supplemental Information S1. Thermal Effects on the spin current nano-oscillator (SCNO) performance as a biosensor**

In the Letter, the performance of the SCNO biosensor was demonstrated at T = 0 K. In Figure S1(a) & (b) the performance of a bare SCNO device has been compared to that of the SCNO biosensor with a single MNP positioned at the center of the device in presence of thermal perturbation. It has been a known fact that for room-temperature performance of magnetoresistive (MR) biosensors, the real-time sensitivity is highly compromised by background noise. Thus, to validate the fact that SCNO devices are better in this respect, we have carried out the theoretical studies at different temperatures ranging from T = 0 K, 60 K, 200 K, 300 K, 400 K and 500 K. The FFT peaks in Figure S1(a) demonstrate that with the gradual increase in temperature, the intensity of the peaks for a bare SCNO device decreases implying the plot becomes noisy at 500K but still the peaks are detectable. Figure S1(b) gives a detailed analysis of the intensity and peak frequency of the bare SCNO biosensor in comparison to that of a SCNO biosensor with a single MNP situated at the center. In both cases, with increase in temperature, the value of peak frequency increases but the intensity decreases significantly. It is evident that at T = 500 K, apart from the intensity (in a.u.) being low, there is no significant shift in frequency for presence of a single MNP ($f_{peak}$ = 3.067 GHz) in comparison to the frequency of a bare SCNO device ($f_{peak}$ = 3.061GHz).

   It is unlikely that the real-time biosensing experiments using SCNO will be conducted at T = 500K. At room temperature (T = 300 K), the difference in peaks for a bare SCNO device and a device with only a single MNP positioned at the center is discernible, both in terms of frequency and intensity (see Figure S1(b); color-symbol code: cyan-stars).

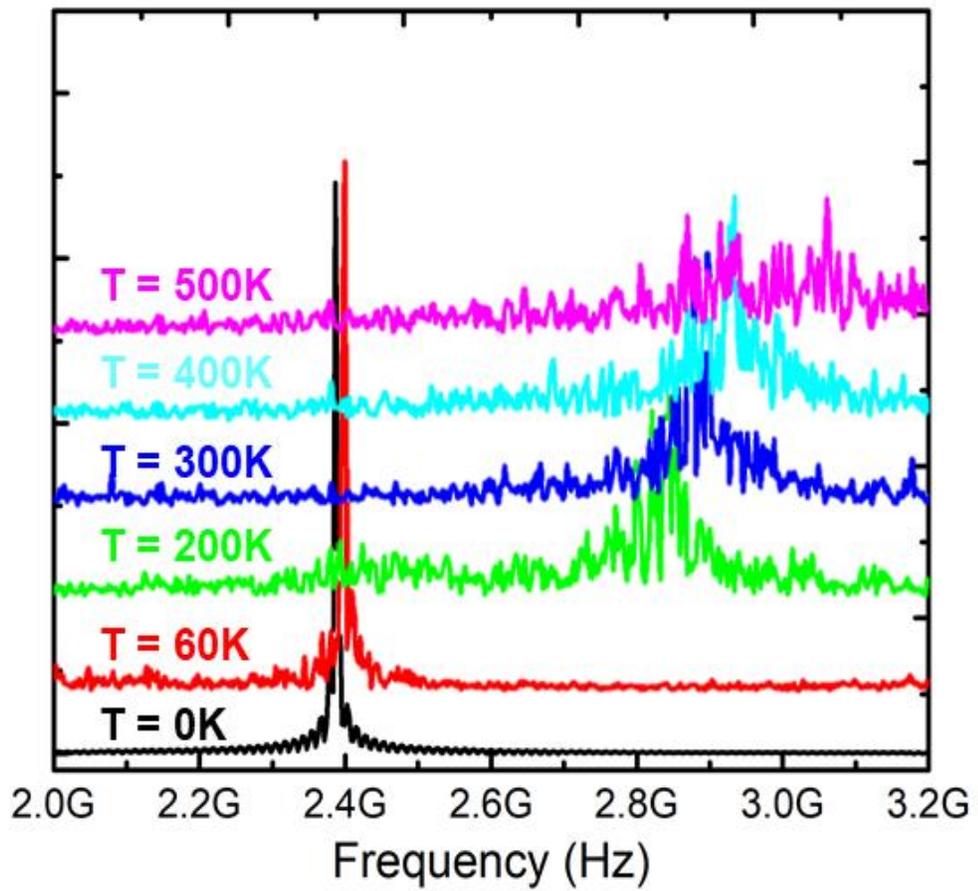

Figure S1(a) The effect of thermal perturbation on the performance of bare spin current nano-oscillator (SCNO) device.

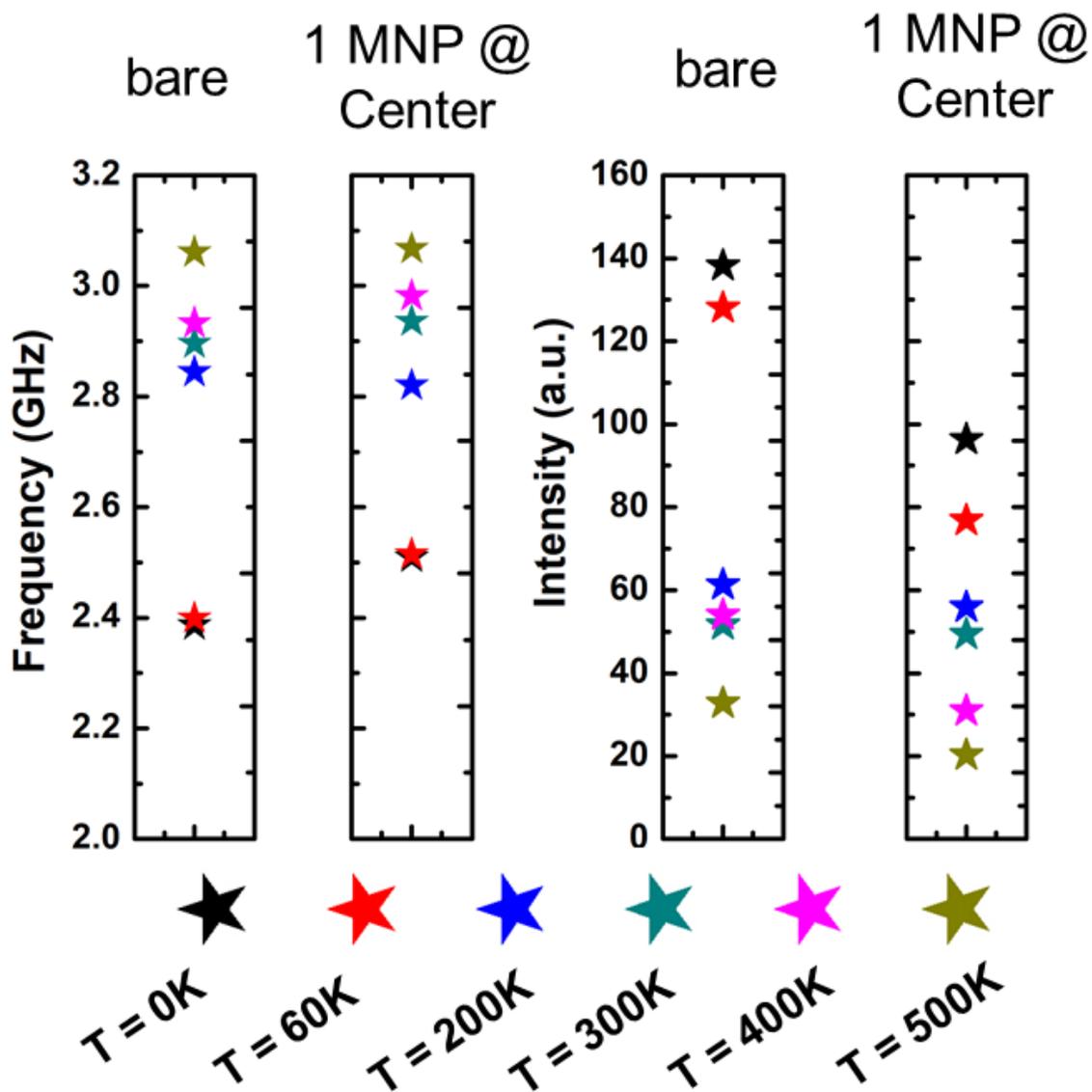

Figure S1(b) Comparison of the peak frequency (in GHz) and intensity (in a.u.) for thermal perturbation between a bare SCNO device and a single MNP situated at the center of the SCNO device.

**Supplemental Information S2. SCNO biosensor performance concerning reversal of current and applied magnetic field direction**

In the Letter, the SCNO device performance has been demonstrated using the current direction to be along **-x** direction and the uniform external DC magnetic field to be directed along **+y** direction. In Figure S2(a)-(c), we observe a case where for a ferromagnetic (FM) nanopillar of dimension 160 nm × 80 nm × 5 nm, the direction of the current is along **+x** direction and the uniform external magnetic field is along **-y** direction. The bare device shows a peak frequency at 2.35 GHz in contrast to 2.387 GHz for the reversed directions. The position dependent performance of the SCNO biosensor remains the same where the cases (ii) & (v) and (iii) & (iv) shows the same peaks (see Figure S2(c)). This too confirms the case for the reversed directions that the two halves of the SCNO biosensor works uniquely.

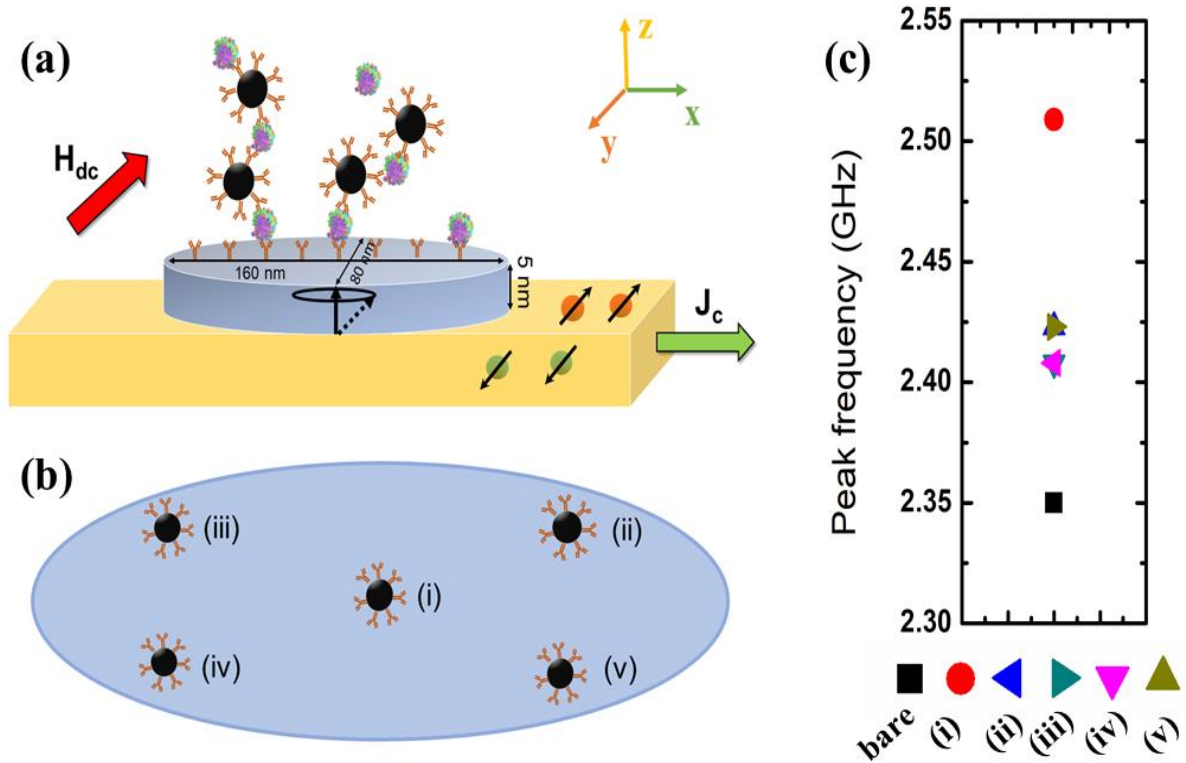

Figure S2. Position dependent sensitivity of the SCNO biosensor with reversed directions of current and applied magnetic field. (a) Schematic of the reversed directions of applied magnetic field and current to the SCNO biosensor. (b) Schematic demonstration of a single MNP on the surface of the biosensor at 5 different positions (i), (ii), (iii), (iv) & (v). Each of the 5 cases are independent of each other. (c) Peak frequency (in GHz) for the 5 different positions of the MNP on the SCNO biosensor with respect to the bare SCNO surface.

**Supplemental Information S3. Magnetic interaction between adjacent ferromagnetic (FM) nanopillars of the SCNO biosensor**

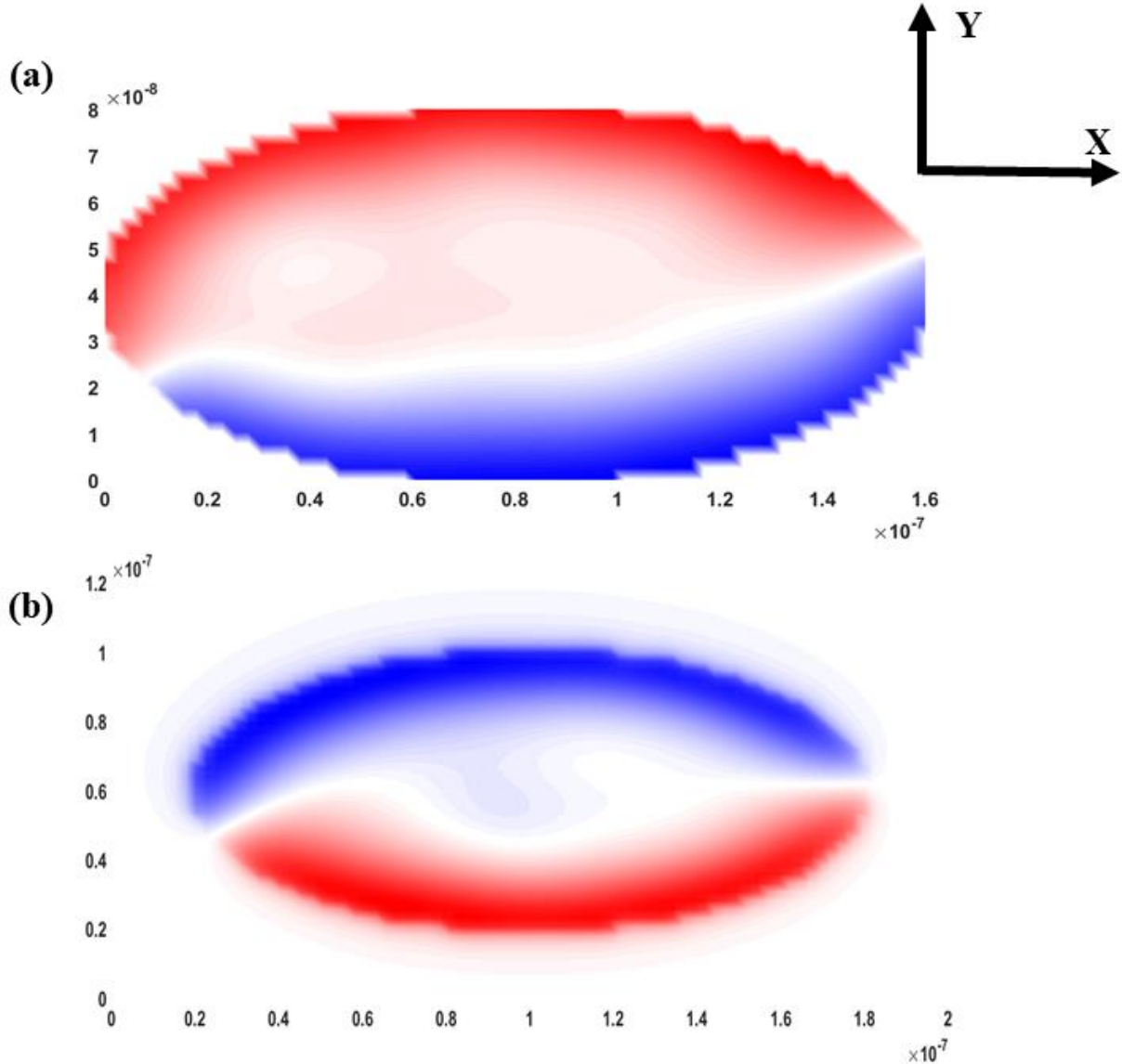

Figure S3. (a) Spatial distribution of the magnetization of the FM nanopillar of dimensions 160 nm × 80 nm × 5 nm when the SCNO was operating in precession mode under a current of 15 mA and a uniform DC magnetic field of 1.1kOe. (b) The spatial distribution of the stray field of the 160 nm × 80 nm × 5 nm FM nanopillar situated exactly at the center of the 200 nm × 120 nm grid space. It is seen that the stray field decays significantly up to a distance of 185 nm along the X axes and up to 120 nm along the Y axes. Therefore, to avoid any interaction between the adjacent FM nanopillars, it is a requirement for them to be situated away from the stray field interference. Hence, we have chosen a safe distance of 0.5μm (for both along X and Y) axes between two adjacent nanopillars (see Figure 1(a)).

**Supplemental Information S4. Comparison of magnetization distribution of FM nanopillar of the SCNO biosensor with and without presence of MNPs**

Figure S4 (a)-(c) represents the magnetization distribution under different conditions of the bare device with or without MNPs when the SCNO device operates in precession mode at a total current of 15 mA along **–x** direction and a uniform DC magnetic field of 1.1 kOe along **+y** direction. The corresponding videos for Figure S4 (a), (b) & (c) are attached in Supplementary Movie SM 1, Supplementary Movie SM 2 and Supplementary Movie SM 3, respectively. The peak frequency for each of the cases (a), (b) & (c) has been reported in the Letter as 2.387 GHz, 2.509 GHz and 3.278GHz respectively (see Figure 3(d) & (f)).

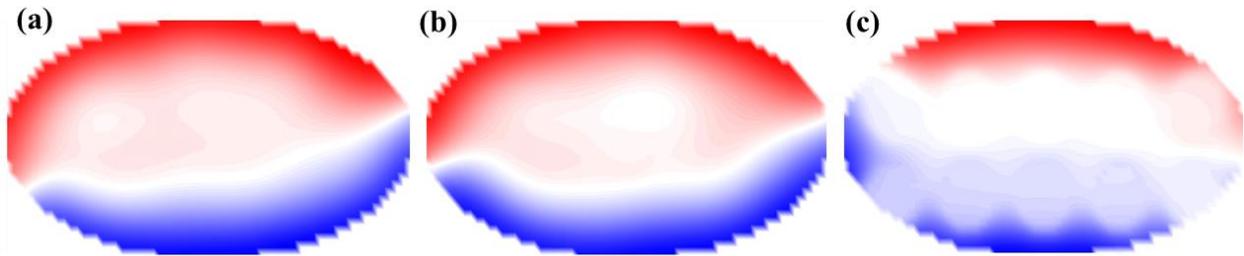

Figure S4. Magnetization distribution of FM nanopillar of the designed 160 nm × 80 nm × 5 nm SCNO biosensor operating in precession mode, under a current of 15 mA and DC magnetic field of 1.1kOe, when (a) bare, (b) only one MNP of size 20 nm present at the center of the device, (c) 10 MNPs, each of size 20 nm, regularly spaced with each of their centers separated by a distance of 30 nm.

**Supplemental Information S5. Power consumption and device performance**

To comment on the power efficiency of the SCNO biosensor and detrimental effect on its sensor performance due to Joule Heating, we have a calculation for the total power consumption and the area overhead in Table S1. The HM/FM layers have been considered to have resistances in parallel. The terms current density (J, Am$^{-2}$) refers to current through the HM only. The total current (I, µA) refers to the current through the entire device. These calculations suggest that SCNO biosensor has a very low power consumption.

Table S1. Power consumption of the SCNO nanopillar

| Parameters | Description | Values |
|---|---|---|
| **FM nanopillar Dimension** | Length × Width × Thickness | 160 nm × 80 nm × 5 nm |
| **HM Dimension** | Length × Width × Thickness | 1µm × 80 nm × 9 nm |
| **J** | Electrical current density | $1.5 \times 10^{12}$ A/m$^2$ |
| **Area of HM** | Width × Thickness | 80 nm × 9 nm |
| **Area of FM** | Width × Thickness | 160 nm × 80 nm × π/4 |
| **ρ$_{HM}$** | Resistivity of HM | $10.6 \times 10^{-8}$ ohm-m |
| **R$_{HM}$** | Resistance of HM | 147.22 ohm |
| **ρ$_{CoFeB}$** | Resistivity of FM | $5.6 \times 10^{-8}$ ohm-m |
| **R$_{CoFeB}$** | Resistance of FM | 4.319 ohm |
| **R$_{eq}$** | Equivalent resistance | 4.196 ohm |
| **I** | Total current | 0.0526 A |
| **P** | Power consumption | 0.0116 W |